\documentclass[aps,pre,twocolumn,amsmath,amssymb,superscriptaddress,floatfix]{revtex4}

\usepackage{graphicx}
\usepackage{bm}
\usepackage[usenames]{color}
\bibstyle{apsrev.bib}

\newcommand{\be}{\begin{equation}}
\newcommand{\ee}{\end{equation}}
\newcommand{\beqn}{\begin{eqnarray}}
\newcommand{\eeqn}{\end{eqnarray}}

\begin{document}

\title{Nonuniversal and anomalous critical behavior of the contact process near an extended defect}
\author{R\'obert Juh\'asz}
\email{juhasz.robert@wigner.mta.hu}
\affiliation{Wigner Research Centre for Physics, Institute for Solid State
Physics and Optics, H-1525 Budapest, P.O.Box 49, Hungary}
\author{Ferenc Igl\'oi}
\email{igloi.ferenc@wigner.mta.hu}
\affiliation{Wigner Research Centre for Physics, Institute for Solid State
Physics and Optics, H-1525 Budapest, P.O.Box 49, Hungary}
\affiliation{Institute of Theoretical Physics, Szeged University, H-6720 Szeged,
Hungary}
\date{\today}

\begin{abstract}
We consider the contact process near an extended surface defect, where the local control parameter deviates from the bulk one by an amount of $\lambda(l)-\lambda(\infty) = A l^{-s}$, $l$ being the distance from the surface. We concentrate on the marginal situation, $s=1/\nu_{\perp}$, where $\nu_{\perp}$ is the critical exponent of the spatial correlation length, and study the local critical properties of the one-dimensional model by Monte Carlo simulations. The system exhibits a rich surface critical behavior. For weaker local activation rates, $A<A_c$, the phase transition is continuous, having an order-parameter critical exponent, which varies continuously with $A$. For stronger local activation rates, $A>A_c$, the phase transition is of mixed order: the surface order parameter is discontinuous, at the same time the temporal correlation length diverges algebraically as the critical point is approached, but with different exponents on the two sides of the transition. The mixed-order transition regime is analogous to that observed recently at a multiple junction and can be explained by the same type of scaling theory.
\end{abstract}


\maketitle

\section{Introduction}
\label{sec:intro}

The contact process \cite{cp,liggett} is the prototype of stochastic lattice models, which undergoes a nonequilibrium phase transition from an active, fluctuating phase to a non-fluctuating (absorbing) one \cite{md,odor,hhl,tauber}. This transition in the homogeneous system belongs to the robust universality class of directed percolation. Although the model is not exactly soluble, the critical exponents and the location of the critical point are known with high precision by series expansions \cite{jd,jensen}. The critical properties of the contact process can be modified by different kind of inhomogeneities. For example near a free surface, where translational invariance is broken, the critical exponent of the local order parameter is different from the bulk one, although the divergence of the correlation length involves the same exponent\cite{jss,egjt,lfh,jensen_s,fhl,hfl}. On the other hand a single defect site represents an irrelevant perturbation, resulting only in correction to scaling terms\cite{bh}. 
Recently, the contact process has been studied near multiple junctions, which are composed of $M>2$ semi-infinite one-dimensional lattices connected to a common central site \cite{mjcp}. Near such a junction the model exhibits a mixed-order transition\cite{bm}: the local order parameter is discontinuous, at the same time the temporal correlation length diverges algebraically as the critical point is approached. The contact process with quenched spatial disorder shows infinite-disorder criticality\cite{moreira,hiv,vojta_rev}, while for temporal disorder, following early numerical works \cite{jensen_td,vazquez}, a so called infinite-noise critical behavior was found by a real-time renormalization group method\cite{vh}. This model has also been studied with long-range interactions \cite{janssen,howard}, on fractals \cite{dsh} and different kinds of complex networks \cite{c_ps,munoz}. We also mention that, in a one-dimensional model of diffusing particles, a single boundary site evolving according to the dynamics of the contact process is able to induce an absorbing phase transition \cite{bh_prl}. 

In this work, we shall consider the contact process in the presence of a smoothly varying inhomogeneity of the power law form, where the local control parameter deviates from the bulk one by an amount 
\be
\Delta \lambda(l)= A l^{-s}\;,
\label{lambda}
\ee
where $l$ measures the distance from the surface. 
In solids, such type of inhomogeneities are known to arise as a result of a uniform elastic deformation \cite{landau_lifshitz}. 
The contact process is often interpreted as a simple, idealised model of population dynamics. In reality, however, it is not uncommon that the local environmental variables which influence the reproduction rate of a population show some variations with the geographical coordinates \cite{gastner}. In this respect, our model describes population dynamics in the presence of a special form of spatial inhomogeneity. According to scaling theory, when lengths are rescaled by a factor $b>1$, so that $l' = l/b$, the inhomogeneity in the local control parameter transforms as: $\Delta \lambda'(l')=b^{1/\nu_{\perp}}\Delta \lambda(l)$, where $\nu_{\perp}$ is the critical exponent of the correlation length measured in the direction of the inhomogeneity. (In systems with isotropic critical scaling there is just one correlation-length exponent, however for the contact process we have different scaling exponents for spatial and temporal correlations.) Therefore the prefactor $A=\Delta \lambda(l)l^s$ obeys the transformation law\cite{burkhardt,cordery}:
\be
A'=A b^{1/\nu_{\perp}-s}\;.
\label{Phi}
\ee
Thus for $s>\nu_{\perp}$, when the decay of the inhomogeneity is fast enough $A$ is decreasing and one expects the same critical behavior as at a clean free surface. On the contrary, for $s<1/\nu_{\perp}$ the decay of the inhomogeneity is so slow, that a new type of singularity of the surface order parameter at the critical point is expected. 

Extended defects of the form in Eq.(\ref{lambda}) have been studied before in two-dimensional Ising\cite{HvL}, Gaussian\cite{burkhardt_guim} and directed walk models\cite{igloi92}, as well as in the mean-field approximation\cite{igloi_palagyi}; for a review, see Ref. \cite{ipt}. According to exact results the local critical behavior in these problems is in agreement with the relevance-irrelevance criterion in Eq.(\ref{Phi}). The most interesting behavior is found for the marginal perturbation, when $s=1/\nu_{\perp}$, in which case the surface critical behavior depends on the parameter $A$ in Eq.(\ref{lambda}). In this paper, we extend these investigations to the one-dimensional contact process, in particular we will study the critical behavior for marginal perturbations through Monte Carlo simulations.
 
The rest of the paper is organised in the following way. In Sec.\ref{Sec:model} the model is defined and details of the numerical calculations are given. Results of the Monte Carlo simulations can be found in Sec.\ref{Sec:results}. Related problems and scaling considerations are presented in Sec.\ref{Sec:related} and our main findings are discussed in Sec. \ref{Sec:discussion}.

\section{The model}
\label{Sec:model}
In the contact process, each site of a lattice can either be vacant ($\bf{\O}$) or be occupied by one particle ($\bf{A}$). The dynamics of the model is a continuous time Markov process in which particles at site $l$ can disappear ($\bf{A} \to \bf{\O}$) with a rate $\mu(l)$, while new particles can be produced on empty sites ($\bf{\O} \to \bf{A}$), with a rate $p\Lambda(l)/n$, where $n$ is the coordination number of the lattice and $p$ is the number of occupied neighbors. Here we consider a one-dimensional semi-infinite lattice, where $l$ measures the distance from the free surface and  the local control parameter defined as $\lambda(l)=\Lambda(l)/\mu(l)$ has a smooth inhomogeneity in the form of Eq.(\ref{lambda}).

In this system, there is a non-equilibrium phase-transition at $\lambda_c=3.29785(2)$\cite{jd}, so that the bulk order-parameter
vanishes as $\rho \sim \Delta^{\beta}$, with a critical exponent $\beta=0.276486(8)$\cite{jensen}, for a small reduced control parameter $\Delta=\lambda-\lambda_c$. At the same time, the correlation lengths in the bulk diverge as: $\xi_{\perp} \sim |\Delta|^{-\nu_{\perp}}$ and $\xi_{\parallel} \sim |\Delta|^{-\nu_{\parallel}}$, with $\nu_{\perp}=1.096854(4)$\cite{jensen} and $\nu_{\parallel}=1.733847(6)$\cite{jensen}, in the spatial and temporal directions, respectively. In the semi-infinite geometry the order parameter at the surface, $\rho_1$, generally shows different type of singularity at $\lambda=\lambda_c$ than its bulk counterpart. In the homogeneous system with $A=0$ there is a second-order surface transition with a modified order-parameter exponent: $\beta_1=0.73371(2)$\cite{jensen_s}, but the singularity of the correlation lengths involve the same exponents as in the bulk: $\nu_{\parallel,1}=\nu_{\parallel}$ and $\nu_{\perp,1}=\nu_{\perp}$. In the present study we consider a marginal smoothly varying perturbation in Eq.(\ref{lambda}) with $s=1/\nu_{\perp}$ and study numerically how the singularity of $\rho_1$ depends on the value of the parameter $A$.

In the numerical simulations, the above continuous-time process is implemented as follows. The coordinates of active sites are stored in a list, and, in every update step, a site is picked randomly. Then, the chosen site is either made inactive with probability $1/[\lambda(l)+1]$, or, one of its equiprobably chosen neighboring sites is activated with probability  $\lambda(l)/[\lambda(l)+1]$ (provided it was inactive). In case of an end site, the adjacent site is activated with probability $\frac{1}{2}\lambda(l)/[\lambda(l)+1]$ (and with the same probability nothing happens).
Time is increased by $1$ after $N(t)$ updates, where  $N(t)$ is the number of active sites at the beginning of the time step. 
We have performed seed simulations, in which all but the first site was initially inactive, and measured the survival probability $P(t)$, which is the probability that the system has not become trapped in the absorbing state up to time $t$, in typically $10^6$ runs. The long time limit of the survival probability $P(t)$ defined in this way is nothing but the surface order parameter, $\rho_1$, of the model.
The size of the system was chosen sufficiently large so that during the simulations the other end site of the lattice was never activated.

\section{Numerical results}
\label{Sec:results}

\begin{figure}[ht]
\includegraphics[width=9cm]{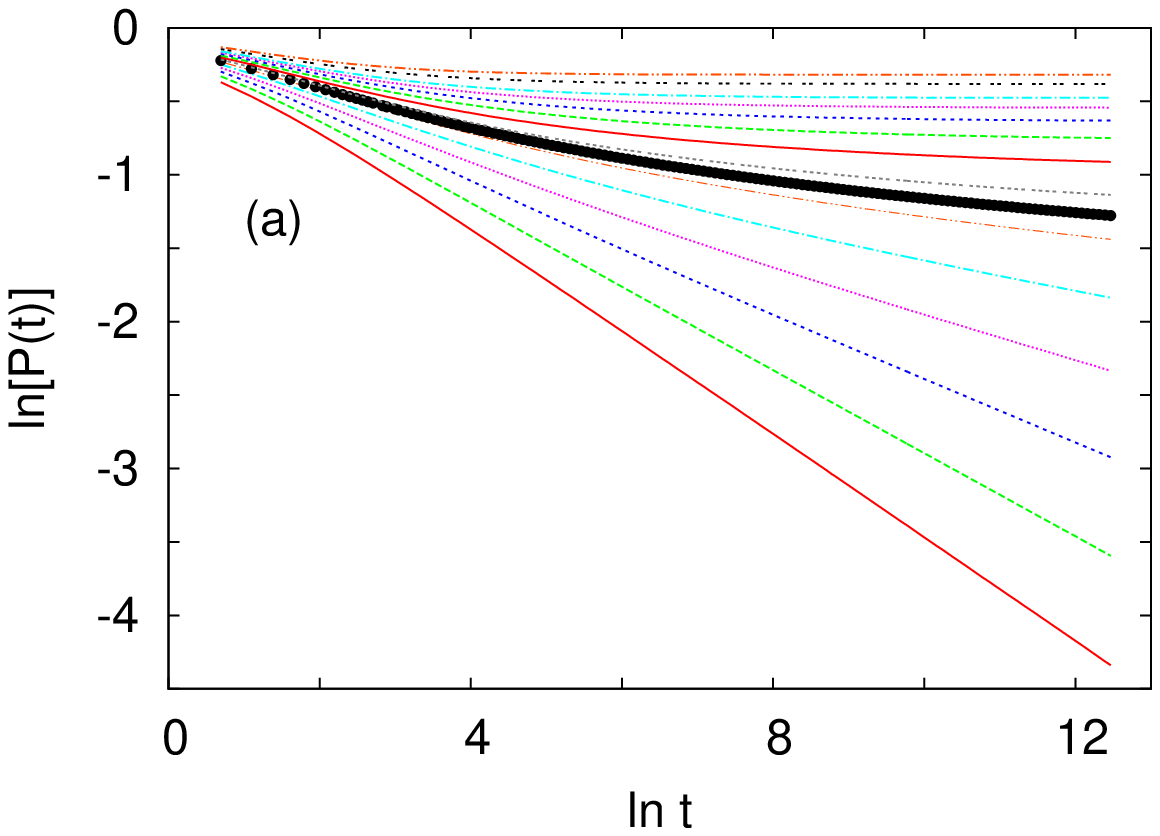} 
\includegraphics[width=9cm]{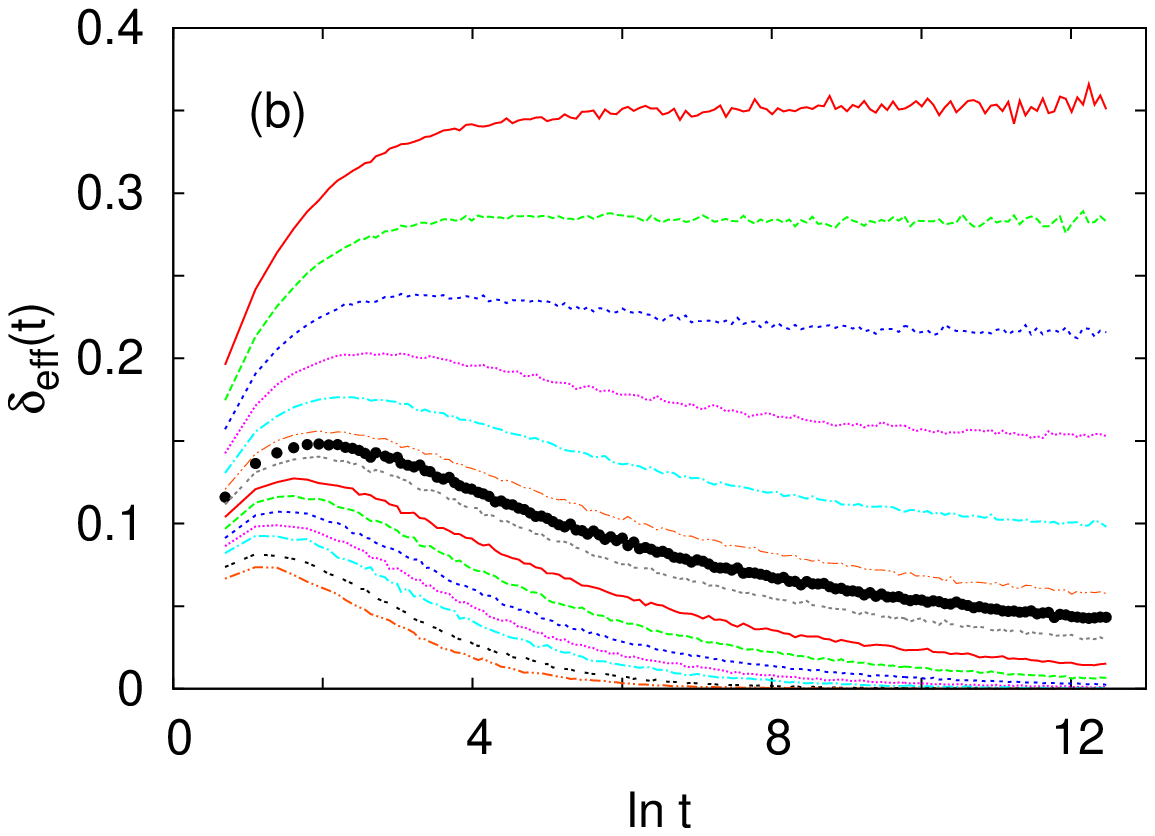} 
\caption{\label{fig_pt} (Color online) (a) Time-dependence of the survival probability measured in numerical simulations for different values of $A$, $A=0.5,1,1.5,2,2.5,3,3.25,3.5,4,4.5,5,5.5,6,7,8$ (from bottom to top) in the bulk critical point $\lambda=\lambda_c$. The data at the estimated tricritical point $A_c=3.25$ are shown by circles. (b) Effective decay exponents for the same data (from top to bottom).}
\end{figure}

\begin{figure}[ht]
\includegraphics[width=9cm]{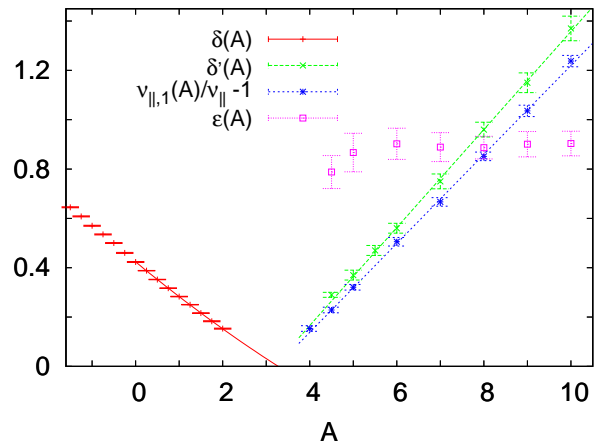}
\caption{\label{exp} (Color online) 
Numerical estimates of various surface critical exponents plotted against $A$. The curves fitted to the data points (quadratic function for $\delta$ and linear for  $\delta^{\prime}$ and $\nu_{\parallel,1}$) seem to cross the x-axis at $A=A_c \approx 3.25$.
}
\end{figure}

\begin{figure}[ht]
\includegraphics[width=9cm]{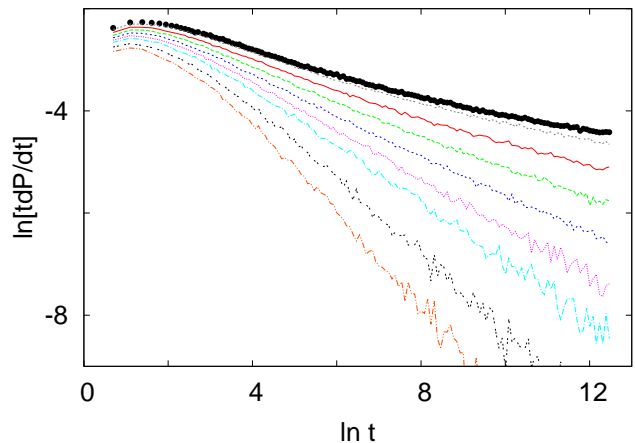}
\caption{\label{fig_der} (Color online) 
Time-dependence of the quantity $tdP(t)/dt$ in a log-log plot for different values of $A$ in the discontinuous regime, 
$A=3.25,3.5,4,4.5,5,5.5,6,7,8$ (from top to bottom). 
The asymptotic slope of the curves is $-\delta^{\prime}$ according to Eq. (\ref{Pt}).
}
\end{figure}

The dependence of the survival probability on time as well as that of the effective decay exponent calculated by 
\be 
\delta_{\rm eff}(t)=-\frac{\ln P(t)-\ln P(t')}{\ln t -\ln t'},
\label{deltaeff}
\ee  
where $t$ and $t'$ are subsequent measuring times (typically $\ln t -\ln t'\approx 0.08$),
are shown in Fig. \ref{fig_pt} in the bulk critical point $\lambda=\lambda_c$ for different values of the parameter $A$. 
As can be seen in the figures, two regimes can be distinguished. 
For small enough $A$, the survival probability decays to zero algebraically, 
\be
P(t)\sim t^{-\delta(A)}, 
\ee
and the decay exponent $\delta(A)$ is a decreasing function of $A$. The estimates of $\delta(A)$ obtained by linear fits to the data presented in Fig. \ref{fig_pt}, are collected in Table \ref{table_I} and are plotted against $A$ in Fig. \ref{exp}. The time-window used in our simulation seems to be not long enough to obtain the asymptotic values of $\delta(A)$ for $A > 2$, in which regime we simply continued analytically the data for $A \le 2$, which is indicated by a full line in Fig. \ref{exp}. This construction shows, that $\delta(A)$ goes to zero at around $A=A_c\approx 3.25$. This observation is in agreement with the behavior of $P(t)$ shown in Fig. \ref{fig_pt}: for $A>A_c$ the survival probability tends to a non-zero limit $p(A)$ for long times thus obviously, $\delta_{\rm eff}(t)$ tends to zero in this regime. In order to find out the way $P(t)$ tends to its limit, we analysed the discrete-time derivative of $P(t)$. The log-log plot shown in Fig. \ref{fig_der} suggests an algebraic approach to the long-time limit in the form 
\be
P(t)\simeq b(A)t^{-\delta'(A)}+p(A),
\label{Pt}
\ee
which can be clearly seen at least for values of $A$ not too close to $A_c$. 
Going closer to $A_c$, the validity of Eq. (\ref{Pt}) is shifted to 
longer times, but presumably it remains valid asymptotically in the entire range $A>A_c$.  


\begin{figure}[ht]
\includegraphics[width=9cm]{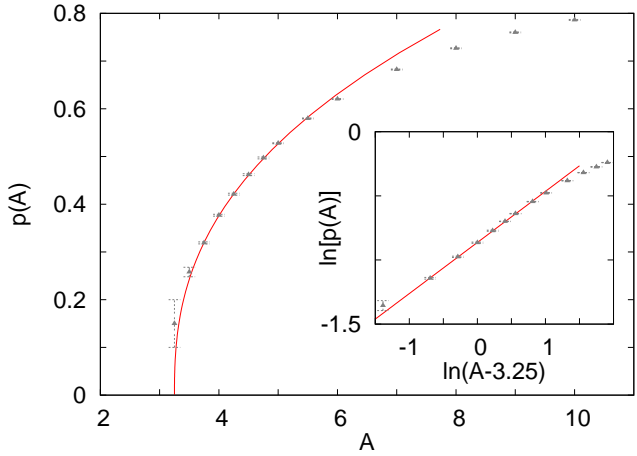} 
\caption{\label{ord} (Color online) Numerically estimated surface order parameters as a function of $A$. The full (red) line represents the extrapolated behavior according to Eq.(\ref{tc}). The slope of the straight line in the inset is $0.40$. 
}
\end{figure}

The surface order parameter $p(A)$ has been estimated by fitting a function given in Eq. (\ref{Pt}) to the putative asymptotic regime of the numerical data. Results can be found in Table \ref{table_II} and seen in Fig. \ref{ord}. 
Using the estimate $A_c=3.25(10)$ for the location of the tricritical point, 
the order parameter is found to vanish according to 
\be
p(A)\sim (A-A_c)^{\beta_{tc}}
\label{tc}
\ee 
as $A_c$ is approached from above, with an exponent $\beta_{tc}=0.40(4)$, see the inset of Fig. \ref{ord}. This asymptotic form fits to the numerical data for $A\ge 3.75$. However, the correction exponent $\delta^{\prime}(A)$ is found to be more sensitive to the size of the time-window and the calculated estimates obtained by linear fits to the data presented in Fig. \ref{fig_der} are stable for $A \ge 4.5$. These are collected in Table \ref{table_II} and are plotted against $A$ in Fig. \ref{exp}. By continuing analytically these data $\delta^{\prime}(A)$ seems to vanish at around $A_c \approx 3.25$ in agreement with the behaviors of $\delta(A)$ and $p(A)$. (It is amusing to notice, that $A_c \approx \lambda_c$, within the error of the calculation.)

\begin{figure}[ht]
\includegraphics[width=9cm]{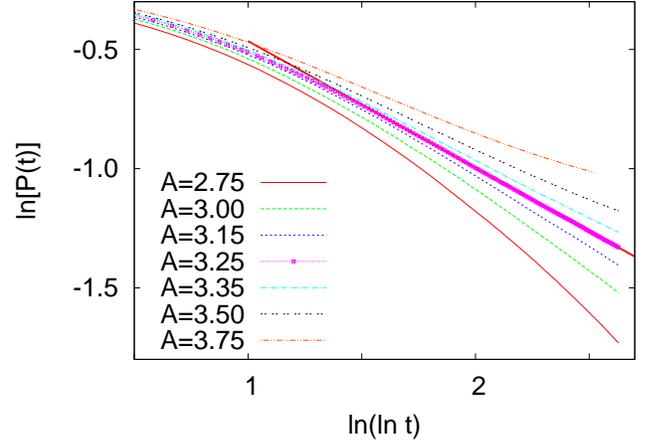} 
\caption{\label{Ploglog} (Color online) 
The logarithm of the survival probability as a function of $\ln(\ln t)$ for different values of $A$. The slope of the straight line is $-0.53$.  
}
\end{figure}

\begin{figure}[ht]
\includegraphics[width=9cm]{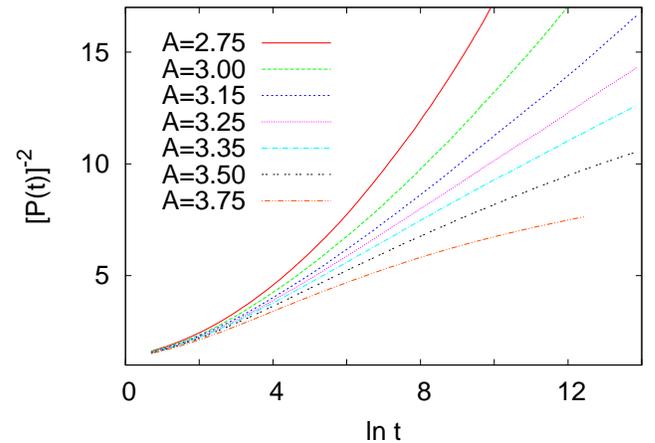}
\caption{\label{inv_log} (Color online) 
The inverse of the square of the survival probability plotted against $\ln t$ for different values of $A$. 
}
\end{figure}

The local critical behavior of the contact process at an extended surface defect seems to be similar to that of the
planar Ising model with an extended surface defect, for which analytical results exist \cite{HvL,bloete_hilhorst,bi,ipt}. In that model, in the marginal case $s=1/\nu=1$, the spatial spin correlation function $G_{\parallel}(r)$ parallel with the surface shows different behaviors depending on $A$. If $A<A_c=1$, $G_{\parallel}(r)$ decays algebraically with the non-universal exponent $2x_s=1-A/A_c$, whereas if $A=A_c$, it decays logarithmically as $G_{\parallel}(r)\sim (\ln r)^{-1}$. In the regime  $A>A_c$, there is a spontaneous surface magnetisation, which vanishes as $(A-A_c)^{1/2}$ as $A_c$ is approached. Here, $G_{\parallel}(r)$ tends to its limiting value according to a power law with the exponent $2x_s'=A/A_c-1$.  
In the contact process, the quantity which is analogous to $G_{\parallel}(r)$ is the density autocorrelation function $C(t_2-t_1)$ in the steady state. The non-stationary scaling behavior of the survival probability is related to the stationary scaling of the autocorrelation function through $C(t)\sim [P(t)]^2$, so in the case of an analogous behavior to that of the Ising model one would expect 
\be 
P(t)\sim (\ln t)^{-\gamma}
\label{Plog}
\ee
in the point $A=A_c$ with some exponent $\gamma$, which may possibly be different from that of the Ising model $\gamma=1/2$. 
Plotting $\ln[P(t)]$ against $\ln(\ln t)$ as shown in Fig. \ref{Ploglog}, we can see that the decay of $P(t)$ is compatible with the logarithmic form in Eq. (\ref{Plog}) at $A_c \approx 3.25$, while, below or above this value, the data curve downward or upward, respectively. A linear fit to the long-time domain gives $\gamma=0.53$, which is quite close to the corresponding value $1/2$ of the Ising model. 
The approximatively logarithmic decay of $P^2(t)$ at $A_c \approx 3.25$ is also demonstrated in Fig. \ref{inv_log}.
We note that using the assumption in Eq.(\ref{Pt}) the estimates from our limited time-series at $A=A_c$ are $p=0.15(5)$ and $\delta_{\rm eff}=0.13(1)$, which illustrates that much longer simulations would be necessary to reach the asymptotic region.


We have also studied the scaling of the survival probability in the case when the bulk is off-critical. 
Let us first consider the regime $A<A_c$, where the surface order parameter exhibits a continuous transition as the bulk control parameter $\Delta\equiv\lambda-\lambda_c$ crosses the critical point $\Delta=0$. 
In the off-critical system, the temporal correlation length $\xi_{\parallel}$ is finite, which manifests itself in the behavior of $P(t)$ either as an exponential cut-off in the inactive phase $\Delta<0$ or an exponential approach to the stationary value in the active phase $\Delta>0$. 
Approaching the critical point from either phase, $\xi_{\parallel}$ diverges according to 
\be 
\xi_{\parallel}\sim |\Delta|^{-\nu_{\parallel,1}(A)}
\ee
and $P(t)$ possesses the scaling property
\be 
P(t,\Delta)=t^{-\delta(A)}f(\Delta t^{1/\nu_{\parallel,1}(A)}),
\label{scaling1}
\ee
where the scaling function $f(x)$ is different in the active and inactive phase.
In accordance with the surface critical behavior of the homogeneous contact process ($A=0$) \cite{hhl}, this scaling relation is fulfilled with the bulk value $\nu_{\parallel,1}(A)=\nu_{\parallel}$ of the correlation-length exponent in the whole domain $A<A_c$, although, in the vicinity of $A_c$, there are strong corrections to scaling.
 
\begin{figure}[ht]
\includegraphics[width=9cm]{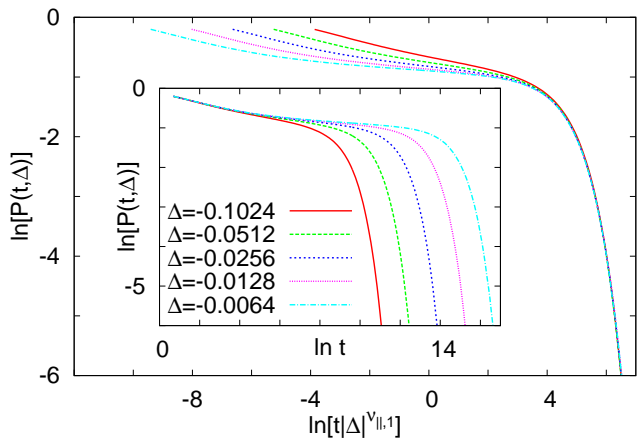} 
\caption{\label{nu4a} (Color online)  
Scaling plot of the survival probability in the inactive phase for $A=4$. The parameter $\nu_{\parallel,1}=2.0$ were used. The inset shows the unscaled data.
}
\end{figure}

Next, let us consider the regime $A>A_c$, where the surface order parameter is non-zero in the bulk critical point $\Delta=0$. 
In the inactive phase, again with strong corrections close to $A_c$, the relation in Eq. (\ref{scaling1}) is found to hold with $\delta(A)=0$ and a correlation-length exponent which depends on $A$, as it is illustrated for $A=4$ in Fig. \ref{nu4a}. The estimates of $\nu_{\parallel,1}(A)$ obtained by achieving a data collapse can be found in Table \ref{table_II} and are plotted in Fig. \ref{exp}. 
Note that the stationary value $p(A)$ has little influence on the quality of the data collapse in Fig. \ref{nu4a}, contrary to the determination of the correction exponent $\delta^{\prime}(A)$. Therefore the estimates of $\nu_{\parallel,1}(A)$ are stable at least in the region $A \ge 4$.

\begin{figure}[ht]
\includegraphics[width=9cm]{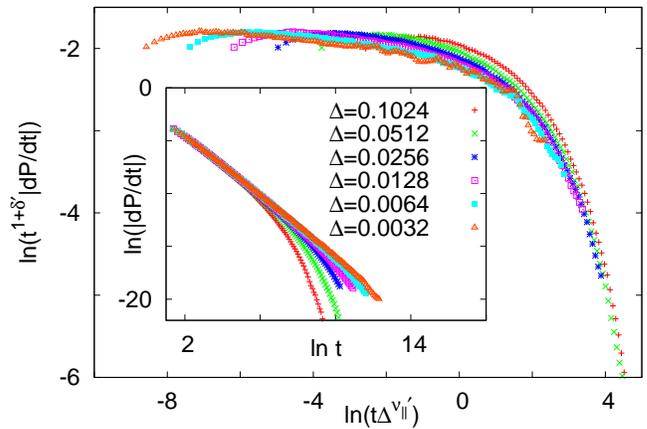} 
\caption{\label{nu5b} (Color online)  
Scaling plot of the derivative of the survival probability in the active phase for $A=5$. The parameters $\delta'=0.37$ and $\nu_{\parallel}^{\prime}=1.7338$ were used. The inset shows the unscaled data.
}
\end{figure}

In the active phase $\Delta>0$, the deviation of $P(t,\Delta)$ from the critical surface order parameter $p\equiv \lim_{t\to\infty}P(t,0)$ is expected to have the scaling property: 
\be 
P(t,\Delta)-p(A)=t^{-\delta^{\prime}(A)}f(\Delta t^{1/\nu_{\parallel}^{\prime}(A)}).
\label{scaling2}
\ee
In order to eliminate $p(A)$, we considered the derivative $dP(t)/dt$ rather than $P(t)$ itself, which behaves according to
\be 
\frac{dP(t,\Delta)}{dt}=t^{-\delta^{\prime}(A)-1}g(\Delta t^{1/\nu_{\parallel}^{\prime}(A)}),
\label{scaling3}
\ee
where $g(x)$ is another scaling function. As it is demonstrated in Fig. \ref{nu5b} for $A=5$, a satisfactory scaling collapse is obtained in accordance with Eq. (\ref{scaling3}) if the bulk correlation-length exponent $\nu_{\parallel}$ is used for $\nu^{\prime}_{\parallel}(A)$. Although the quality of the data collapse seems to be better by using a somewhat lower value, $1.6$, we attribute this deviation to corrections to scaling. 

\begin{table}[h]
\begin{center}
\begin{tabular}{|c|l|}
\hline  $A$  &  $\delta$ \\ \hline
-1.50  & 0.645(2) \\ \hline
-1.25  & 0.608(2) \\ \hline
-1.00  & 0.570(2) \\ \hline
-0.75  & 0.535(2) \\ \hline
-0.50  & 0.500(2) \\ \hline
-0.25  & 0.460(2) \\ \hline
0.00   & 0.42317(1) \\ \hline
0.25   & 0.388(1) \\ \hline
0.50   & 0.352(1) \\ \hline
0.75   & 0.317(1) \\ \hline
1.00   & 0.283(1) \\ \hline
1.25   & 0.250(1) \\ \hline  
1.50   & 0.216(1) \\ \hline
1.75   & 0.183(2) \\ \hline   
2.00   & 0.153(2) \\ \hline   
\end{tabular}
\end{center}
\caption{\label{table_I} Estimated decay exponents for different values of $A$ in the continuous regime. The value for the homogeneous system ($A=0$) is taken from Ref. \cite{jensen_s}.}
\end{table}

\begin{table}[h]
\begin{center}
\begin{tabular}{|c|l|l|l|}
\hline  $A$  &  $\delta'$ & $\nu_{\parallel,1}$ & $p$ \\ \hline
3.25 &   &             &       0.15(5) \\ \hline
3.5  &   &             &       0.26(1) \\ \hline
3.75 &   &             &       0.319(2) \\ \hline
4.0  &   &   2.00(2)   &       0.377(2) \\ \hline
4.25 &   &             &       0.421(2) \\ \hline
4.5  & 0.29(1)  &   2.13(2)   &       0.462(2) \\ \hline
5.0  & 0.37(2)  &   2.29(2)   &       0.528(1) \\ \hline
5.5  & 0.47(2)  &             &       0.580(1)   \\ \hline
6.0  & 0.56(2)  &   2.61(3)   &       0.620(1) \\ \hline
7.0  & 0.75(3)  &   2.89(3)   &       0.682(1) \\ \hline
8.0  & 0.96(3)  &   3.21(3)   &       0.727(1)  \\ \hline
9.0  & 1.15(4)  &   3.53(4)   &       0.760(1) \\ \hline
10.0 & 1.37(5)  &   3.88(4)   &       0.786(1) \\ \hline
\end{tabular}
\end{center}
\caption{\label{table_II} Estimated values of the correction exponent $\delta'$, the correlation-length exponent $\nu_{\parallel,1}$, and the order parameter $p$ for different values of $A$.}
\end{table}

\section{Related problems and scaling considerations}
\label{Sec:related}

In the previous section, we have presented numerical results about the local critical behavior of the contact process at an extended, marginal surface defect. Here we compare these results with the local critical behavior in similar problems. First, we consider the local critical behavior at a multiple junction and use this analogy to present a scaling theory, which explains some of our numerical results. Then we consider the case when the extended defect is placed in the bulk of the system. Finally, we present some scaling results about relevant extended surface defects.

\subsection{Critical behavior at multiple junctions}

A multiple junction is such a geometry, in which $M>2$ semi-infinite chains are connected to a central site\cite{indekeu,itb,mjcp,grassberger}. The local critical behavior of the contact process at multiple junctions has recently been studied and mixed-order transition has been observed\cite{mjcp}. The critical exponents at the junction, such as $\delta'$ and $\nu_{\parallel,1}$ in the inactive phase are $M$-dependent, and comparing their values with the series of data in Table \ref{table_II}, we can assign an approximate effective parameter $A_M$ for each value of $M$. These are $A_{3} \approx 4.75$, $A_{4} \approx 7.3$ and $A_{5} \approx 9.5$. The observed critical behavior at multiple junctions has been explained within the frame of a scaling theory in Refs.\cite{it,mjcp}, which is expected to hold for the problem of the marginal extended defect in the region $A>A_c$. In the following we recapitulate the essential points of this scaling reasoning.

\subsection{Scaling consideration for $A>A_c$}

We start to write the survival probability (the surface order parameter) $P=P(\Delta,h_1,\Delta_1,t)$ as a function of the bulk and surface control parameters, $\Delta$ and $\Delta_1$, respectively, as well as the surface ordering field, $h_1$. According to the scaling theory in Ref.\cite{mjcp}, when lengths are rescaled by a factor $b>1$, so that $l \to l/b$, the survival probability satisfies the relation:
\be
P(\Delta,h_1,\Delta_1,t)=b^{-z+y_{h_1}}\tilde{P}(\Delta b^{1/\nu_{\perp}},h_1 b^{y_{h_1}},\Delta_1 b^{y_{\Delta_1}},t/b^z)\;.
\label{rho_scal1}
\ee
Here $y_{h_1}$ and $y_{\Delta_1}$ are the scaling exponents associated to $h_1$ and $\Delta_1$, respectively and $z=\nu_{\parallel}/\nu_{\perp}$ is the dynamical exponent. For $A>A_c$ the surface is ordered at the transition point, thus $P=p$ is constant and scale independent, consequently $y_{h_1}=z$.

The essence of the scaling theory, that we assume, is that the irrelevant variable $\Delta_1$ has different properties in the active and in the inactive phases, respectively. In the active phase, $\Delta \ge 0$, $\Delta_1$ is a \textit{harmless} variable, thus the scaling functions are analytic and can be expanded in Taylor series:
\be
P(\Delta,h_1,\Delta_1,t)-p= b^{y_{\Delta_1}} \hat{P}(\Delta b^{1/\nu_{\perp}},h_1 b^{y_{h_1}},t/b^z)\;.
\label{rho_scal2}
\ee
From this equation, using standard scaling considerations, we obtain the exponent relations: $\delta'=-y_{\Delta_1}/z$ and $\beta'=-y_{\Delta_1} \nu_{\perp}$.

In the inactive phase, $\Delta < 0$, $\Delta_1$ is a \textit{dangerous} variable and the scaling functions are nonanalytic in $\Delta_1$. For the temporal correlation length we assume the following functional form:
\be
\xi_{\parallel}(\Delta,h_1,\Delta_1,l)=\Delta_1^{-\epsilon}\xi_{\parallel}(\Delta,h_1\Delta_1^{-\epsilon},l)\;,
\label{xi_par}
\ee
from which it follows, that the correlation-length exponent at the surface is given by: $\nu_{\parallel,1}=\nu_{\parallel}(1+\epsilon \delta')$. From this relation we can extract the new exponent:
\be
\epsilon=(\nu_{\parallel,1}/\nu_{\parallel}-1)/\delta'\;,
\label{epsilon}
\ee
what has been calculated from the measured data in Table \ref{table_II}. The results are presented in Fig.\ref{exp}. For large enough values of $A$, $A > 4$, the calculated exponents are saturated around $\epsilon \approx 0.9$, the same value that has been obtained for multiple junctions\cite{mjcp}. In the region $A_c<A<4$, we have no accurate estimates for the critical exponents in the large time limit, but presumably $\epsilon$ remains constant all the way to $A_c$. 

\subsection{Extended defect in the bulk}

The extended defect can also be placed in the bulk, so that the inhomogeneity in the local control parameter assumes the form: 
$\lambda(l)-\lambda = A |l|^{-s}$, where $-\infty<l<\infty$ now measures the distance from the center of the defect. The relevance-irrelevance criterion in this case is the same as for the surface defect, see in Eq.(\ref{Phi}). Having the marginal perturbation, $s=\nu_{\perp}$, according to numerical investigations now the critical value is at $A_c=0$. Note that the same is true for the planar Ising model\cite{ibt}. For $A \le 0$, the phase transition at the defect is continuous and the local critical exponents are $A$-dependent. For small enough value of $A <-2$ the local critical exponents are approximately the same as in the semi-infinite problem. Thus the extended defect seems to act as an effective cut. For negative values of $A$ which are closer to $0$, the calculated effective exponents have not reached their asymptotic value within the available time-window, but have a tendency to continuously increase with $t$. The measured effective exponents at the largest available time are smaller, than in the semi-infinite system. Whether this trend stays valid also for the asymptotic values, we can not decide due to strong corrections to scaling. For $A>0$ the phase transition at the defect is of mixed order and the corresponding critical exponents are $A$-dependent again.

We have also checked the local critical behavior at multiple junctions with $M>2$ having also a marginal extended defect. In this case, the two regimes (non-universal continuous transition for $A \le A_c$ and mixed-order transition for $A>A_c$) are separated by a tricritical point at $A_c<0$. For example with $M=3$, the tricritical point is estimated as $A_c \approx -0.7$.

\subsection{Scaling considerations for relevant inhomogeneities}

For slowly varying inhomogeneities with $s<1/\nu_{\perp}$ the perturbation is relevant and a new type of surface critical behavior is induced. The singular behavior of the surface order parameter can be calculated through a scaling theory\cite{ipt}. For enhanced surface couplings, $A>0$, the survival probability (surface order parameter) stays finite at the bulk critical point and it satisfies the scaling relation:
\be
P(\Delta,A)=b^{-\beta_1/\nu_{\perp}} P(\Delta b^{1/\nu_{\perp}},A b^{1/\nu_{\perp}-s}),\quad A>0\;,
\ee
where we have used the scaling law of $A$ in Eq.(\ref{Phi}).
Now putting $b=\xi_{\perp}$ we obtain:
\be
P(\Delta,A)=\Delta^{\beta_1} f\left(\frac{\ell}{\xi_{\perp}}\right),\quad \ell=|A|^{-\nu_{\perp}/(1-\nu_{\perp}s)}\;,
\label{ell}
\ee
where $\ell$ is the (finite) length-scale induced by the inhomogeneity. The scaling function $f(x)$ for small argument behaves as: $f(x) \sim x^{\omega}$ with $\omega=-\beta_1/\nu_{\perp}$, which ensures that the $\Delta$-dependence cancels, thus $P$ stays constant. Consequently, the survival probability at the bulk critical point behaves as:
\be
p \sim \ell^{\omega} \sim |A|^{\beta_1/(1-\nu_{\perp}s)},\quad A>0\;.
\ee
The spatial correlation function between two sites, $G(l_1,l_2)$, can be calculated by noticing that in the vicinity of the bulk transition point a smoothly varying local length scale can be defined as:
\be
\xi_{\perp}(l) \sim \left[\Delta \lambda(l)\right]^{-\nu_{\perp}}\;,
\ee
and the complete correlation function is obtained in a form of an integral:
\beqn
G(l_1,l_2) &\sim& \exp\left( -\int_{l_1}^{l_2} \frac{{\rm d} l}{\xi_{\perp}(l)}\right) \cr
&\sim& \exp\left[-a\left(l_2^{1-s\nu_{\perp}} -l_1^{1-s\nu_{\perp}}\right)\right]\;.
\label{G}
\eeqn
Thus the spatial correlations at the bulk transition point between the surface and the bulk are given in a stretched exponential form:
\be
G(0,l) \sim \exp\left[-\left(l/\ell\right)^{1-s\nu_{\perp}}\right]\;,
\label{G1}
\ee
where $\ell$ is defined in Eq.(\ref{ell}).

For reduced surface couplings ($A<0$) in the active phase ($\Delta>0$), the surface order is induced by the bulk order parameter at a distance $r$ from the surface. $r$ can be estimated by equating the contributions from the bulk control parameter, $\Delta$, and that of the inhomogeneity $|A|r^{-s}$, leading to $r \sim (|A|/\Delta)^{1/s}$. Now we argue, that the order parameter near the surface is proportional to the correlation function at $r$, thus we obtain from Eq(\ref{G1}):
\be
P(\Delta,A) \sim \exp\left[-a |A|^{1/s} \Delta^{\nu_{\perp}-1/s}\right]\;.
\ee

\section{Discussion}
\label{Sec:discussion}
In this paper, we have studied the surface critical behavior of the one-dimensional contact process in the presence of a smooth inhomogeneity of the power-law form in Eq.(\ref{lambda}) and have concentrated on the marginal perturbation with $s=\nu_{\perp}$. In this case, for weak surface couplings having a parameter $A<A_c$, the surface phase transition is continuous and the critical exponent of the surface order parameter is a continuous function of $A$. At the same time, the correlation-length exponents, $\nu_{\perp}$ and $\nu_{\parallel}$ stay constant and retain their bulk values. Interestingly, we can tune the parameter to such a value, $A \approx 1.95 \approx 0.6 A_c$, that the critical behavior at the surface is identical to that in the bulk. For strong surface couplings with $A>A_c$, the surface transition is of mixed order: the surface order parameter is discontinuous at the transition point, while the correlation lengths are divergent. In this regime, the temporal correlation length shows an exponent asymmetry: in the active phase the estimate is compatible with the bulk value, while in the inactive phase it exceeds the bulk value and increases with $A$. The jump in the surface order parameter vanishes algebraically as the tricritical point is approached and at the tricritical point $A=A_c$, the survival probability exhibits a logarithmic decay.

We have shown that the critical behavior in the mixed-order transition regime ($A>A_c$) is similar to that at a multiple junction having $M>2$ legs. The critical behavior of both problems can be described by the same type of scaling theory, in which an irrelevant parameter plays important r\^ole. One can define a combination of the measured exponents, $\epsilon=(\nu_{\parallel,1}/\nu_{\parallel}-1)/\delta'$ which turns out to be (approximately) independent of the parameters $A$ and $M$, having the same value $\epsilon \approx 0.9$. We note that similar observation has been made for the planar Ising model, in which case this exponent is exactly known as $\epsilon=1$\cite{mjcp}.

If the extended defect is placed in the bulk, then the continuous and mixed-order transition regimes are separated at $A_c=0$ and for (large enough) negative values of $A$, the defect acts as an effective cut.

Finally, we mention that the extended defect problem can be generalised to higher dimensions, as well. 
The defect can then be located either at the surface along which it is translationally invariant or centered in the bulk so that the system is translationally invariant in a subspace of $d_{\rm def}<d$ dimensions and rotationally symmetric in the orthogonal subspace of $d-d_{\rm def}$ dimensions.  
In the case of a surface defect or a bulk defect with $d_{\rm def}=d-1$, a scenario similar to that of the one-dimensional problem is expected to hold. 
If, however,  $d_{\rm def}<d-1$, the local critical behavior can be different. 
For example, for a rotationally symmetric bulk defect with $d_{\rm def}=0$ in $d=2$ dimensions, no local ordering at the center of the defect is expected, in analogy with the similar problem for the planar Ising model\cite{ipt,radial}.

\begin{acknowledgments}
We thank P. Grassberger, G. \'Odor, and L. Turban for useful comments. 
This work was supported by the Hungarian Scientific Research Fund under Grants No. K109577 and No. K115959.
\end{acknowledgments}


\end{document}